\documentclass[12pt]{article}
\begin{document}
\title{Generalized Hamiltonian structures for Ermakov
systems}
\author{F.~Haas  \\
Laborat\'orio Nacional de Computa\c{c}\~ao Cient\'{\i}fica \\
Coordena\c{c}\~ao de Matem\'atica Aplicada  \\
Av. Get\'ulio Vargas, 333 \\
25651-070, Petr\'opolis, RJ - Brazil\\
ferhaas@lncc.br}
\date{\relax}
\maketitle
\begin{abstract} \noindent
We construct Poisson structures for Ermakov systems, using the
Ermakov invariant as the Hamiltonian. Two classes of Poisson
structures are obtained, one of them degenerate, in which case we
derive  the Casimir functions. In some situations, the existence
of Casimir functions can give rise to superintegrable Ermakov
systems. Finally, we  characterize the cases where linearization
of the equations of motion is possible.
\end{abstract}


\section{Introduction}

Ermakov systems \cite{Ermakov}-\cite{L67} have attracted attention
in the last decades due to their important physical applications
and mathematical properties as well. A central mathematical
property of Ermakov systems is the existence of a constant of
motion, the Ermakov invariant. The Ermakov invariant allows to
construct a nonlinear superposition law linking the solutions of
the equations of motion composing the Ermakov system \cite{Reid}.
Ermakov systems have recently been of interest in diverse
scenarios, such as accelerator physics \cite{Drivotin},
dielectric planar waveguides \cite{Rosu1}, cosmological models
\cite{Rosu2, Bertoni}, analysis of supersymmetric families of
Newtonian free damping modes \cite{Rosu3}, study of open fermionic
systems \cite{Kim}, analysis of the propagation of electromagnetic
waves in one-dimensional inhomogeneous media \cite{guasti},
algebraic approach to integrability of nonlinear systems
\cite{band}, coupled linear oscillators \cite{thy}, the
semiclassical limit of quantum mechanics \cite{Matzkin},
supersymmetric quantum mechanics \cite{kaushal}, computation of
geometrical angles and phases for nonlinear systems
\cite{mama1}-\cite{mama3}, search for Noether \cite{Simic, Haas1}
and Lie \cite{Leach, Haas2} symmetries, the possible linearization
of the system \cite{Athorne1, Haas3}, extension of the Ermakov
system concept \cite{Leach}, \cite{Bassom}-\cite{Atho}, the search
for additional constants of motion \cite{Goedert} and some
discretizations of Ermakov systems \cite{Michelin, Hone}.

The existence of a Hamiltonian or Lagrangian formulation is a
central question for any dynamical system. Cerver\'o and Lejarreta
\cite{Cervero} have identified a Hamiltonian subclass of Ermakov
systems and have used this Hamiltonian formulation as the starting
point for the quantization of these systems. Latter on, Haas and
Goedert have extended the class of Hamiltonian Ermakov systems by
inclusion of frequency functions depending not only on time, but
on dynamical variables as well \cite{Haas4}. Both Hamiltonian
formulations for Ermakov systems are canonical formulations, for
which the Poisson bracket is defined in the conventional way. On
the other hand, non canonical, or generalized Hamiltonian, or
Poisson descriptions, have proven to be relevant in such diverse
fields like magnetohydrodynamics, kinetic models in plasma
physics, biological models, optics, quantum chromodynamics and so
on \cite{Bermejo1}. There is a wide range of possibilities open
when a Poisson formulation is available, such as nonlinear
stability analysis through the energy-Casimir method, perturbation
methods, integrability results and bifurcation properties
\cite{Bermejo1}. Accordingly, in recent years there has been
interest in constructing Poisson structures
\cite{Gumral}-\cite{Haas5}, mainly for the special case of
three-dimensional dynamical systems.

A finite-dimensional dynamical system is said to be a generalized
Hamiltonian system when it can be cast in the form
\begin{equation} \label{eq4} \dot{x}^\mu =
J^{\mu\nu}\partial_{\nu}H \,, \quad \mu = 1,\dots,N \,,
\end{equation}
where sum convention is assumed and $\partial_\mu =
\partial/\partial x^{\mu}$. Here, $H = H(x)$ is the Hamiltonian
function and $J^{\mu\nu} = J^{\mu\nu}(x)$ is the Poisson matrix
for the system. The Poisson matrix must be skew-symmetric,
$J^{\mu\nu} = - J^{\nu\mu}$. Moreover, it must satisfy the
following set of partial differential equations,
\begin{equation}
\label{jacobi} J^{\mu\nu}\partial_{\nu}J^{\rho\sigma} +
J^{\rho\nu}\partial_{\nu}J^{\sigma\mu} +
J^{\sigma\nu}\partial_{\nu}J^{\mu\rho} = 0 \,.
\end{equation}
These equations ensure  that the  generalized Poisson bracket,
defined as
\begin{equation}
\label{pb} \{A,B\} = \partial_{\mu}A J^{\mu\nu} \partial_{\nu}B
\,,
\end{equation}
for any functions $A = A(x)$ and $B = B(x)$,  satisfies the Jacobi
identity
\begin{equation}
\{A,\{B,C\}\} + \{B,\{C,A\}\} + \{C,\{A,B\}\} = 0 \,.
\end{equation}
In fact, equations (\ref{jacobi}) are a necessary and sufficient
condition for the bracket (\ref{pb}) to satisfy the Jacobi
identity. Hereafter we refer to (\ref{jacobi}) as the ``Jacobi
identities'' too. The above generalized Poisson brackets are
endowed with all properties of conventional Poisson brackets, with
the advantage of being applicable to more general systems.

From the definition, we can identify the basic building blocks of any Poisson
formulation as being the Hamiltonian function and the Poisson
matrix. If a time-independent constant of motion is known, the
idea is trying to use it as the Hamiltonian function for the
system. In other words,  we can look (\ref{eq4}) in a reverse way,
as a set of equations for some of the components of the Poisson
matrix. Then, for given $\dot{x}^\mu$ and $H$,   system
(\ref{eq4}) is an undetermined linear system for the matrix
elements $J^{\mu\nu}$. Besides being skew symmetric, the Poisson
matrix must comply with the Jacobi identities (\ref{jacobi}), that
constitute an overdetermined system of partial differential
equations for the remaining components $J^{\mu\nu}$ not fixed by
(\ref{eq4}). This ``deductive schema'' for constructing Poisson
structures was developed in detail in the reference \cite{Haas5},
and applied to three-dimensional dynamical systems like
Lotka-Volterra systems for three interacting populations, the
Rabinovich system and the Rikitake dynamo model \cite{Goedert2}.
The basic proposition of the present work is to put forward the
approach of ref. \cite{Haas5}, to find new classes of Ermakov
systems for which a Hamiltonian formalism is possible. We use the
only time-independent constant of motion always available for
Ermakov systems, namely the Ermakov invariant, as the Hamiltonian
function and check the consequences of this assumption. This idea
is partly inspired by the results of reference \cite{ath}, where
it was shown that $(n+1)-$dimensional extensions of Ermakov
systems, when restricted to the unit sphere $S^n$, can sometimes
be viewed as canonical Hamiltonian systems with the Ermakov
invariant playing the role of Hamiltonian. Here, however, there is
no restriction to any particular submanifold, and we focus on
non-canonical descriptions. Finally, notice that Ermakov systems
are non autonomous four-dimensional dynamical systems, in
contradistinction to the earlier works \cite{Goedert2, Haas5}
focused on three-dimensional models.

The paper is organized as follows. In Section 2, we set the
Hamiltonian for Ermakov systems as being the Ermakov invariant,
and seek a Poisson matrix complying with the Jacobi identities. In
this way, we arrive at two classes of Ermakov systems admitting
Poisson structures, analyzed in detail in Section 3. In Section 4,
we examine the possibility of applying linearization transforms to
the resulting Ermakov systems with Poisson character. Section 5 is
dedicated to our final remarks.

\section{Poisson structures}

Classical Ermakov systems are commonly written in the form
\begin{eqnarray}
\ddot x + \omega^{2} x &=& \frac{1}{yx^2}f(y/x) \,, \\
\ddot y + \omega^{2} y &=& \frac{1}{xy^2}g(x/y) \,,
\end{eqnarray}
where $f$ and $g$ are arbitrary functions of the indicated
variables, and $\omega$ is an arbitrary frequency function. In
most applications, $\omega$ is restricted to be dependent on time
only, but here this constraint is relaxed.

For our purposes, polar coordinates $r = (x^2 + y^2)^{1/2}$ and
$\theta = \arctan(y/x)$ are more appropriate. Ermakov systems in
polar coordinates reads
\begin{eqnarray}
\label{eq1} \ddot{r} - r{\dot\theta}^2 + \omega^{2}\,r &=&
\frac{1}{r^3}F(\theta) \,,
\\ \label{eq2} r\ddot\theta + 2\dot{r}\dot\theta &=& - \frac{1}{r^3}G(\theta) \,,
\end{eqnarray}
where $F$ and $G$, depending only on the angle variable, are
arbitrary functions suitably related to $f$ and $g$. The main
property of Ermakov systems is that they always possess a constant
of motion, the Ermakov invariant,
\begin{equation} \label{eq3} I = \frac{1}{2}(r^{2}\dot\theta)^2 +
\int^{\theta}G(\lambda)d\lambda \,. \end{equation}
As manifest by (\ref{eq3}), the existence of $I$ is not affected
by any particular dependence on $\omega$ of the dynamical
variables. Also, a little thought shows that for frequency
functions depending on dynamical variables we can set $F \equiv 0$
in equation (\ref{eq1}) without any loss of generality. However,
we keep $F$ mainly for ease comparison with previous results on
Ermakov systems.

We want to reformulate our Ermakov system (\ref{eq1}-\ref{eq2}) as
a generalized Hamiltonian system as in (\ref{eq4}), with all
indices running from $1$ to $4$ since the Ermakov system is a
system of two second order ordinary differential equations. The
Poisson matrix $J^{\mu\nu}$ is skew symmetric and should satisfy
the following system of partial differential equations,
\begin{eqnarray}
\label{jac1} J^{\mu\,1}\partial_{\mu}J^{23} +
J^{\mu\,2}\partial_{\mu}J^{31} + J^{\mu\,3}\partial_{\mu}J^{12}
&=& 0 \,, \\ \label{jac2} J^{\mu\,1}\partial_{\mu}J^{24} +
J^{\mu\,2}\partial_{\mu}J^{41} + J^{\mu\,4}\partial_{\mu}J^{12}
&=& 0 \,, \\ \label{jac3} J^{\mu\,1}\partial_{\mu}J^{34} +
J^{\mu\,3}\partial_{\mu}J^{41} + J^{\mu\,4}\partial_{\mu}J^{13}
&=& 0 \,, \\ \label{jac4} J^{\mu\,2}\partial_{\mu}J^{34} +
J^{\mu\,3}\partial_{\mu}J^{42} + J^{\mu\,4}\partial_{\mu}J^{23}
&=& 0 \,. \end{eqnarray}

If time-independent, the Hamiltonian $H$ is a constant of motion.
The only time-independent constant of motion always available for
Ermakov systems, no matter are the functions $F$, $G$ and
$\omega$, is the Ermakov invariant. Hence it is natural to define
\begin{equation} \label{eq5} H = I \end{equation}
and see the consequences.

The choice of coordinates $x^\mu$ is a matter of convenience.
Here, we choose
\begin{equation}
\label{eq6} (x^{1},x^{2},x^{3},x^{4}) = (r,\theta,u,v) \,,
\end{equation}
where
\begin{equation}
\label{eq7} u = \dot{r} \,, \quad v = r^{2}\dot\theta \,.
\end{equation}
The Ermakov system viewed as a first order system then reads
\begin{eqnarray} \label{eq8} \dot r
&=& u \,,\\ \label{eq9} \dot\theta &=& v/r^2 \,,\\ \label{eq10}
\dot u &=& - \omega^{2}r + (v^2 + F(\theta))/r^3 \,,\\
\label{eq11} \dot v &=& - G(\theta)/r^2 \,. \end{eqnarray}

Using the Ermakov invariant as the Hamiltonian,
\begin{equation} \label{eq12} H = \frac{v^2}{2} +
\int^{\theta}G(\lambda)d\lambda \,, \end{equation}
there results, from (\ref{eq4}) and (\ref{eq8}-\ref{eq10}), that
\begin{eqnarray} \label{eq13} u
&=& J^{12}G(\theta) + J^{14}v \,, \\ \label{eq14} v/r^2 &=&
J^{24}v \,, \\ \label{eq15} - \omega^{2}r + (v^2 + F(\theta))/r^3
&=& J^{32}G(\theta) + J^{34}v \,. \end{eqnarray}
Equation (\ref{eq11}) follows automatically from $\dot{H} = 0$ and
the skew symmetry of $J^{\mu\nu}$ (see \cite{Haas5} for details).

Let us look more closely the system (\ref{eq13}-\ref{eq15}).
Equation (\ref{eq15}) can be viewed as the definition of $\omega$,
while equation (\ref{eq13}) shows that setting $J^{12} = 0$
eliminates $G(\theta)$ from all considerations.

This is a convenient choice, and still leads to a large class of
examples. We found, after long calculations, that is very hard to
impose the Jacobi identities when $J^{12} \neq 0$. Thus, in what
follows we set $J^{12} = 0$, leaving $G(\theta)$ arbitrary and
allowing for more general classes of Ermakov systems.

Summing up results from (\ref{eq13}-\ref{eq15}) and our choice for
$J^{12}$, we obtain
\begin{eqnarray} \label{eq16} J^{12} &=& 0 \,,\\ \label{eq17} J^{14} &=& u/v \,, \\ \label{eq18}
J^{24} &=& 1/r^2 \,,\\ \label{eq19} \omega^2 &=& (v^2 +
F(\theta))/r^4 + (J^{23}G(\theta) - J^{34}v)/r \,. \end{eqnarray}
Our goal now is to insert (\ref{eq16}-\ref{eq18}) into Jacobi
identities and solve for the remaining components of the Poisson
matrix. With the solution, we can know what are the allowable
frequencies using (\ref{eq19}).

The second Jacobi identity, equation (\ref{jac2}), gives
\begin{equation} \label{eq20}
J^{23} = \frac{u}{r^{2}v} \,. \end{equation}
Inserting this and (\ref{eq16}-\ref{eq18}) into (\ref{jac1}), we
obtain
\begin{equation} \label{eq21} u\frac{\partial J^{13}}{\partial u} +
v\frac{\partial J^{13}}{\partial v} = J^{13} - \frac{u^2}{v^2} \,,
\end{equation}
with solution
\begin{equation} \label{eq22} J^{13} = \left(\frac{u}{v}\right)^2 + u\psi\left(\frac{u}{v},r,\theta,t\right) \,.
\end{equation}
Here, $\psi$ is an arbitrary function of the indicated arguments.
Notice that we have included time-dependence for extra generality.

Substituting the already   calculated components of the Poisson
matrix into (\ref{jac4}) gives
\begin{equation}
\label{eq23} u\frac{\partial J^{34}}{\partial u} + v\frac{\partial
J^{34}}{\partial v} = J^{34} + \frac{2uv}{r}\psi \,,
\end{equation}
whose solution is
\begin{equation}
\label{eq24} J^{34} =
\frac{2uv}{r}\psi\left(\frac{u}{v},r,\theta,t\right) +
u\varphi\left(\frac{u}{v},r,\theta,t\right) \,,
\end{equation}
where $\varphi$ is a further arbitrary function, of the indicated
arguments.

The only Jacobi identity still deserving attention is the equation
(\ref{jac3}), which yields the consistency condition
\begin{equation}
\label{eq25} \psi\varphi' - \psi'\varphi =
\frac{\partial\psi}{\partial r} +
\frac{v}{r^{2}u}\frac{\partial\psi}{\partial\theta} -
\frac{2}{r}\psi \,,
\end{equation}
where the prime denotes differentiation with respect to $u/v$. For
$\psi = 0$, the consistency condition (\ref{eq25}) is satisfied in
an immediate way leaving $\varphi$ arbitrary. For $\psi \neq 0$, a
different class of solutions arises.

These two possibilities and the associated Poisson structures are
studied separately.

\section{The two classes of solution}

\subsection{The case $\psi = 0$.}

For $\psi = 0$, the consistency condition (\ref{eq25}) imposes no
constraints on the function  $\varphi$, which remains arbitrary.
From the results of the last section, we obtain the following
Poisson matrix,
\begin{equation}
\label{eq26} J^{\mu\nu} = \pmatrix{0&0&(u/v)^2&u/v\cr
0&0&u/(r^{2}v)&1/r^2\cr -(u/v)^2&-u/(r^{2}v)&0&u\varphi\cr
-u/v&-1/r^{2}&-u\varphi&0\cr} \,,
\end{equation}
where, as said previously, $\varphi = \varphi(u/v,r,\theta,t)$. By
construction,  this is a Poisson matrix. Moreover, it is not of
Lie-Poisson, affine-linear or quadratic type, as more usual
\cite{Bermejo1}.

The frequency function of the associated Ermakov system follows
from (\ref{eq19}),
\begin{equation}
\label{y1} \omega^2 = \frac{1}{r^4}(v^2 + F(\theta)) +
\frac{u}{r^{3}v}G(\theta) - \frac{uv}{r}\varphi(u/v,r,\theta,t)
\,.
\end{equation}
Using the frequency function as defined in (\ref{y1}) and the
definitions of $u$ and $v$ in terms of the original polar
coordinates, we derive the following Ermakov system,
\begin{eqnarray}
\label{y2} \ddot r &=& - \frac{\dot r}{r^{4}\dot\theta}G(\theta) +
r^{2}\dot{r}\dot\theta\varphi\left(\frac{\dot r}{r^{2}\dot\theta},r,\theta,t\right) \,, \\
\label{y3} r\ddot\theta + 2\dot{r}\dot\theta &=& -
\frac{G(\theta)}{r^3} \,.
\end{eqnarray}
We see that the function $F$ disappears from all considerations.
By construction, (\ref{y2}-\ref{y3}) is an Ermakov system having a
Poisson formulation, with the Hamiltonian being the Ermakov
invariant and the Poisson matrix given by (\ref{eq26}). This is an
infinite family of Ermakov systems, containing two arbitrary
functions, $G$ and $\varphi$. Also notice that the frequencies as
given by (\ref{y1}) can not be functions of time only, necessarily
having a dependence on the dynamical variables.

Here, the Poisson structure is degenerate, as can be readily seen
by
\begin{equation}
\label{eq27} \det(J^{\mu\nu}) = 0 \,.
\end{equation}
Therefore, we can obtain Casimir functions, that is, functions
which Poisson commute with any function defined on phase space.
The defining equations for the Casimir functions, denoted by $C$,
are
\begin{equation}
\label{eq28} J^{\mu\nu}\partial_{\nu}C = 0 \,.
\end{equation}
The existence of non constant solutions is due to the degenerate
character of the Poisson structure.

There are systematic methods \cite{Bermejo3, Bermejo4} for
obtaining the Casimirs, but here a direct approach is sufficient.
Using the Poisson matrix, we find the following equations for the
Casimirs,
\begin{eqnarray}
\label{eq29}
u\frac{\partial C}{\partial u} + v\frac{\partial C}{\partial v} &=& 0 \,, \\
\label{eq30} \frac{1}{r^2}\frac{\partial C}{\partial\theta} +
\frac{u}{v}\frac{\partial C}{\partial r} + u\varphi\frac{\partial
C}{\partial u} &=& 0 \,.
\end{eqnarray}
The first equation of this system shows that $C$ do depends only
on the variables $(u/v,r,\theta,t)$. Taking into consideration
this information, we transform the equation (\ref{eq30}) into
\begin{equation}
\label{eq31} \frac{1}{r^2}\frac{\partial C}{\partial\theta} +
\alpha\frac{\partial C}{\partial r} +
\alpha\varphi(\alpha,r,\theta,t)\frac{\partial C}{\partial\alpha}
= 0 \,,
\end{equation}
where $\alpha = u/v$. The solution for (\ref{eq31}) strongly
depends on the details of the function $\varphi$. Notice that
$G(\theta)$, the extra arbitrary function defining the Ermakov
system, does not play any role it the computation of the Casimirs.

An illuminating way to rewrite (\ref{eq31}) is found by means of
the change of coordinates
\begin{equation}
\bar{r} = 1/r \,, \quad \bar\theta = \theta \,, \quad \bar\alpha =
- \alpha \,.
\end{equation}
The equation for the Casimirs becomes
\begin{equation}
\label{y4} \frac{\partial C}{\partial\bar\theta} +
\bar{\alpha}\frac{\partial C}{\partial\bar{r}} +
\frac{\bar\alpha}{\bar{r}^2}\varphi(-\bar\alpha,1/\bar{r},\bar\theta,t)\frac{\partial
C}{\partial\bar\alpha} = 0 \,.
\end{equation}
For $\bar\theta$ interpreted as the independent variable,
$\bar{r}$ as coordinate and $\bar\alpha$ as velocity, this is
Liouville's equation for the invariants of the equations of motion
\begin{equation}
\label{y5} \frac{d\bar{r}}{d\bar\theta} = \bar\alpha \,, \quad
\frac{d\bar\alpha}{d\bar\theta} =
\frac{\bar\alpha}{\bar{r}^2}\varphi(-\bar\alpha,1/\bar{r},\theta,t)
\,,
\end{equation}
which are also the characteristic equations for (\ref{y4}). Notice
that here the physical time $t$ is a mere parameter.

Equations (\ref{y5}) are equivalent to the Newton equation for
one-\-di\-men\-sio\-nal motion under the force field
$\bar\alpha\varphi/\bar{r}^2$. For functions $\varphi$ yielding
completely integrable examples of such motions, we can find all
the Casimirs for the Poisson structure (\ref{eq26}). To show a
concrete example where this is possible, consider the case
\begin{equation}
\label{y8} \varphi = - \frac{\bar{r}^2}{\bar\alpha}
\frac{dV}{d\bar{r}}(\bar{r},t) \,,
\end{equation}
for an arbitrary pseudo potential $V(\bar{r},t)$. In this case,
Newton's equation that follows from (\ref{y5}) are
\begin{equation}
\label{sing} \frac{d^{2}\bar{r}}{d\bar\theta^2} = -
\frac{dV}{d\bar{r}}(\bar{r},t) \,,
\end{equation}
an autonomous potential system, since $\bar\theta$ does not appear
explicitly. As for any autonomous one-dimensional potential
system, there is complete integrability. The constants of motion
are
\begin{eqnarray}
\label{y6}
C_1 &=& \frac{1}{2}\left(\frac{d\bar{r}}{d\bar\theta}\right)^2 + V(\bar{r},t) \,, \\
\label{y7} C_2 &=& \bar\theta -
\frac{1}{\sqrt{2}}\int^{\bar{r}}\frac{d\lambda}{(C_1 -
V(\lambda,t))^{1/2}} \,,
\end{eqnarray}
respectively the energy and the additional integration constant
for the equations of motion.  In terms of the original
coordinates of the Poisson description, the quantities
(\ref{y6}-\ref{y7}) are
\begin{eqnarray}
\label{k1} C_1 &=& \frac{1}{2}\left(\frac{u}{v}\right)^2 +
V(1/r,t) \,, \\ \label{k2} C_2 &=& \theta -
\frac{1}{\sqrt{2}}\int^{1/r}\frac{d\lambda}{(C_1 -
V(\lambda,t))^{1/2}} \,,
\end{eqnarray}
which are the Casimirs for the Poisson structure (\ref{eq26}) when
$\varphi$ is given as in (\ref{y8}). When $V$ does not contain
explicitly the time, the Casimirs becomes additional constants of
motion for the Ermakov system. In this case, we obtain a
superintegrable \cite{Evans} class of Ermakov systems, possessing
three invariants, namely the Ermakov invariant and the two Casimir
functions. In fact, we can derive superintegrable Ermakov systems
in all cases when (\ref{y4}) can be solved in closed form for the
two Casimirs of the Poisson structure, and when $\varphi$ is
time-independent.

Finally, let us examine more closely the superintegrable Ermakov
systems with the Casimirs $C_1$ and $C_2$ given in
(\ref{k1}-\ref{k2}), in the special situation for which $\partial
V/\partial t = 0$. Using $C_2$ as in (\ref{k2}) and the implicit
function theorem, we obtain locally the equation for the orbits,
$r = r(\theta, C_{1}, C_{2})$. Now, using the Ermakov invariant,
we get the angle as a function of time through the quadrature
\begin{equation}
\label{ttt} t + k = \int^{\theta}\frac{r^{2}(\lambda, C_{1},
C_{2}) d\lambda}{h(\lambda, I)} \,,
\end{equation}
where $k$ is the last integration constant and
\begin{equation}
\label{hhh} h(\theta,I) = \sqrt{2}\left(I - \int^{\theta}
G(\lambda)d\lambda\right)^{1/2} \,.
\end{equation}
Locally, (\ref{hhh}) gives $\theta$ as a function of time and four
integration constants, namely $I$, $C_1$, $C_2$ and $k$.

To show an example of the procedure, consider the particular case
\begin{equation}
\label{vvv} V(\bar{r}) = \frac{1}{2\bar{r}^2} \,,
\end{equation}
for which (\ref{sing}) describes a singular oscillator. Using
(\ref{y8}), there results
\begin{equation}
\varphi = - \frac{r^{3}\dot\theta}{\dot r}
\end{equation}
and, from (\ref{y2}-\ref{y3}), we obtain the following Ermakov
system,
\begin{eqnarray}
\ddot r &=& - \frac{\dot{r}}{r^{4}\dot\theta} G(\theta) - r^{5}{\dot\theta}^2 \,, \\
r\ddot\theta + 2\dot{r}\dot\theta &=& - \frac{1}{r^3}G(\theta) \,.
\end{eqnarray}
We left the function $G(\theta)$ undetermined.  From the pseudo
potential (\ref{vvv}) and the orbit equation (\ref{k2}), we get
\begin{equation}
r^2 = \frac{2C_1}{1 + 4C_{1}^{2}(\theta - C_{2})^2} \,,
\end{equation}
showing a spiral motion of a particle coming arbitrarily close to
the origin. The time-dependence of such motion is obtained from
the quadrature (\ref{ttt}), which depends on the details of the
function $G(\theta)$.

\subsection{The case $\psi \neq 0$.}

For $\psi \neq 0$, the consistency condition (\ref{eq25}) has a
different class of solutions,
\begin{eqnarray}
\label{x1} \varphi &=&
\Bigl(\int^{u/v}\frac{d\lambda}{\psi^{2}(\lambda,r,\theta,t)}\Bigl(\frac{\partial\psi}{\partial
r}(\lambda,r,\theta,t) + \frac{1}{r^{2}\lambda}\frac{\partial\psi}{\partial\theta}(\lambda,r,\theta,t) \nonumber \\
&-& \frac{2}{r}\psi(\lambda,r,\theta,t)\Bigr) +
\chi(r,\theta,t)\Bigr) \psi(u/v,r,\theta,t) \,,\end{eqnarray}
where $\chi$ is an arbitrary function of the indicated arguments.
Therefore, we obtain the Poisson structure
\begin{equation}
\label{x2} J^{\mu\nu} = \pmatrix{0&0&(u/v)^2 + u\psi&u/v\cr
0&0&u/(r^{2}v)&1/r^2\cr -(u/v)^2 - u\psi&-u/(r^{2}v)&0&u\varphi +
2uv\psi/r\cr -u/v&-1/r^{2}&-u\varphi - 2uv\psi/r &0\cr} \,,
\end{equation}
with $\varphi$ specified in terms of $\chi$ and $\psi$ according
to (\ref{x1}). The Poisson structure   is non degenerate,
\begin{equation}
\det(J^{\mu\nu}) = \frac{u^{2}\psi}{r^4}(\frac{2u}{v^2} + \psi)
\neq 0 \,.
\end{equation}

The following frequency functions are derived from (\ref{eq19}),
\begin{equation}
\label{x3} \omega^2 = \frac{1}{r^4}(v^2 + F(\theta)) +
\frac{u}{r^{3}v}G(\theta) -
\frac{uv}{r}\left(\varphi(u/v,r,\theta,t) +
2\frac{v}{r}\psi(u/v,r,\theta,t)\right) \,.
\end{equation}
again necessarily depending on the dynamical variables. The
resulting Ermakov systems are
\begin{eqnarray}
\label{x4} \ddot r &=& - \frac{\dot r}{r^{4}\dot\theta}G(\theta) +
r^{2}\dot{r}\dot\theta\left(\varphi(\frac{\dot
r}{r^{2}\dot\theta},r,\theta,t) +
2r\dot\theta\psi(\frac{\dot r}{r^{2}\dot\theta},r,\theta,t)\right) \,, \\
\label{x5} r\ddot\theta &+& 2\dot{r}\dot\theta = -
\frac{G(\theta)}{r^3} \,.
\end{eqnarray}
These Ermakov systems contains three arbitrary functions, namely
$G$, $\psi$ and $\chi$. Since the Poisson structure is non
degenerate, there are no nontrivial Casimirs. Therefore, we found
a new class of Ermakov systems that can be cast in a non
degenerate generalized Hamiltonian form.

\section{Linearization}

Ermakov systems with frequency functions depending only on time
have shown to be linearizable \cite{Athorne1} by use of
\begin{equation}
\label{a1} \bar{r} = 1/r
\end{equation}
as the dependent variable and the angle $\theta$ as the
independent one. This change of variables is accomplished by the
relation
\begin{equation}
\label{a2} \dot\theta = h(\theta,I)/r^2 \,,
\end{equation}
where $h(\theta,I)$ is defined in (\ref{hhh}). In terms of
$\bar{r}$, $\theta$, the Ermakov systems transforms into an
one-parameter family of second-order linear ordinary differential
equations depending on the value of the Ermakov invariant
\cite{Athorne1}, whenever the frequency function does not contain
dynamical variables. This result was used for the analysis of the
stability and periodicity of some Ermakov systems arising in
two-layer, shallow water wave theory \cite{athor}. The
linearization transform (\ref{a1}-\ref{a2}) was shown to be useful
also for a class of Ermakov systems for which the frequency
function is not a mere function of time \cite{Haas3}. This result
provides an explanation for the reasons why Kepler-Ermakov systems
\cite{att}, a perturbation of conventional Ermakov systems, are
linearizable.

In view of the usefulness of the linearization transform
(\ref{a1}-\ref{a2}) in several contexts, it is interesting to
check if it can be also possible for our classes of Ermakov
systems admitting Poisson formulations. For instance, applying
(\ref{a1}-\ref{a2}) to the Ermakov system (\ref{y2}-\ref{y3}), the
result is
\begin{equation}
\label{a3} \frac{d^{2}\bar{r}}{d\theta^2} =
\frac{1}{\bar{r}^2}\frac{d\bar{r}}{d\theta}
\varphi\left(-\frac{d\bar{r}}{d\theta},\frac{1}{\bar{r}},\theta,t\right)
\,.
\end{equation}
For $\partial\varphi/\partial t \neq 0$, (\ref{a3}) becomes an
integro-differential equation, a possibility we will not consider
here. Equation (\ref{a3}) is equivalent to the equation (\ref{y5})
determining the Casimirs for the Poisson structure.

The term on the right hand side of (\ref{a3}) has linear character
if and only if
\begin{equation}
\label{a4} \frac{1}{\bar{r}^2}\frac{d\bar{r}}{d\theta}
\varphi\left(-\frac{d\bar{r}}{d\theta},\frac{1}{\bar{r}},\theta,t\right)
= A(\theta)\frac{d\bar{r}}{d\theta} + B(\theta)\bar{r} + C(\theta)
\,,
\end{equation}
for functions $A$, $B$ and $C$ depending only on the angle. If the
assumption (\ref{a4}) is satisfied, the Ermakov systems
(\ref{y2}-\ref{y3}) falls in the class of linearizable Ermakov
systems discussed in \cite{Haas3}. Moreover, if (\ref{a4}) is
valid the characteristic equations (\ref{y5}) for the Casimirs are
also linear. This does not imply, of course, that the Casimirs may
always be found in closed form when (\ref{a4}) holds. Similar
remarks apply to the linearization of our second class of Ermakov
systems admitting a Poisson formulation, treated in subsection
3.2.

\section{Conclusion}

In this paper we have proposed the Ermakov invariant as the
Hamiltonian function and we reformulate the Ermakov system as a
Poisson system. The main difficulty is to find a Poisson matrix
reproducing the equations of motion and, at the same time, being
compatible with the Jacobi identities. However, the task can be
achieved, and two classes of Poisson structures were derived. One
of them is degenerate, thus opening the possibility of
constructing Casimir invariants. These Casimirs, if
time-independent, are also constants of motion. In the cases where
the Casimirs are time-independent and available in closed form, we
obtain superintegrable Ermakov systems. A class of such
superintegrable Ermakov systems was explicitly shown in Section
3.1. Another, non degenerate, class of Ermakov systems admitting
generalized Hamiltonian formulation with the Ermakov invariant
playing the role of Hamiltonian function was also found. In this
latter case, no Casimirs exist. Both classes of Ermakov systems
are specified by two arbitrary functions. All these considerations
apply to frequency functions having a dependence on the dynamical
variables. Finally, the possibility of linearization of the
equations of motion was analyzed in Section IV. Interestingly, we
found that the determining equations for the Casimirs are linear
when the associated Ermakov systems are linearizable through
(\ref{a1}-\ref{a2}).

\vskip 1cm \noindent{\bf Acknowledgements}\\  This work has been
supported by the Brazilian agency Conselho Nacional de
Desenvolvimento Cient\'{\i}fico e Tecnol\'ogico (CNPq).


\begin{thebibliography}{99}
\bibitem{Ermakov} Ermakov V. P. 1880 {\it Univ. Isv. Kiev} {\bf 20} 1.
\bibitem{RR1} Ray J. R. and Reid J. L. 1979 {\it Phys. Lett. A} {\bf 71} 317.
\bibitem{L67} Lewis H. R. 1967 {\it Phys. Rev. Lett.} {\bf 18} 510.
\bibitem{Reid} Reid J. L. and Ray J. R. 1980 {\it J. Math. Phys.} {\bf 21} 1583.
\bibitem{Drivotin} Drivotin J. I. and Ovsyannikov D. A 1999 {\it Proceedings of the 1999
Particle Accelerator Conference}, New York.
\bibitem{Rosu1} Rosu H. and Romero J. L. 1999 {\it Nuovo Cimento B} {\bf 114} 569.
\bibitem{Rosu2} Rosu H, Espinoza P. and Reyes M. 1999 {\it Nuovo Cimento B} {\bf 114} 1435.
\bibitem{Bertoni} Bertoni C., Finelli F. and Venturi G. 1998 {\it Phys. Lett. A} {\bf 237} 331.
\bibitem{Rosu3} Rosu H. and Espinoza P. 2001 {\it Phys. Rev. E} {\bf 63} 037603.
\bibitem{Kim} Kim S. P., Santana A. E. and Khanna F. C. 2000 {\it Phys. Lett. A} {\bf 272} 46.
\bibitem{guasti} Guasti F., Diamant R. and Villegas A. G. 2000 {\it Rev. Mexicana Fis.} {\bf 46} 530.
\bibitem{band} Bandyopadhyay J. N., Lakshminarayan A. and Sheorey V. B. 2001 {\it Phys. Rev. A} {\bf 63} 042109.
\bibitem{thy} Thylwe K-E. and Korsch H. J. 1998 {\it J. Phys. A: Math. Gen.} {\bf 31} L279.
\bibitem{Matzkin} Matzkin A. 2001 {\it J. Phys. A: Math. Gen.} {\bf 34} 7833.
\bibitem{kaushal} Kaushal R. S. and Parashar D. 1996 {\it J. Phys. A: Math. Gen.} {\bf 29} 889.
\bibitem{mama1} Maamache M. 1995 {\it Phys. Rev. A} {\bf 52} 936.
\bibitem{mama2} Maamache M. 1996 {\it J. Phys. A: Math. Gen.} {\bf 29} 2833.
\bibitem{mama3} Maamache M. 1997 {\it Ann. of Phys. (NY)} {\bf 254} 1.
\bibitem{Simic} Simic S. S. 2000 {\it J. Phys. A: Math. Gen.} {\bf 33} 5435.
\bibitem{Haas1} Haas F. and Goedert J. 2001 {\it Phys. Lett. A} {\bf 279} 181.
\bibitem{Leach} Leach P. G. L. 1991 {\it Phys. Lett. A} {\bf 158} 102.
\bibitem{Haas2} Goedert J. and Haas F. 1998 {\it Phys. Lett. A} {\bf 239} 348.
\bibitem{Athorne1} Athorne C., Rogers C., Ramgulam U. and Osbaldestin A. 1990 {\it Phys. Lett.
A} {\bf 143} 207.
\bibitem{Haas3} Haas F. and Goedert J. 1999 {\it J. Phys. A: Math. Gen.} {\bf 32} 2835.
\bibitem{Bassom} Schief W. K., Rogers C. and Bassom A. P. 1996 {\it J. Phys. A: Math.
Gen.} {\bf 29} 903.
\bibitem{rog} Rogers C. and Schief W. K. 1996 {\it J. Math. Anal. Appl.} {\bf 198} 194.
\bibitem{Atho} Athorne C. 1999 {\it J. Math. Anal. Appl.} {\bf 233} 552.
\bibitem{Goedert} Goedert J. 1989 {\it Phys. Lett. A} {\bf 136} 391.
\bibitem{Michelin} Common A. K. and Musette M. 1997 {\it Phys. Lett. A} {\bf 235} 574.
\bibitem{Hone} Hone A. N. W. 1999 {\it Phys. Lett. A} {\bf 263} 347.
\bibitem{Cervero} Cerver\'o J. M. and Lejarreta J. D. 1991 {\it Phys. Lett. A} {\bf 156} 201.
\bibitem{Haas4} Haas F. and Goedert J. 1996 {\it J. Phys. A: Math. Gen.} {\bf 29} 4083.
\bibitem{Bermejo1} Hern\'andez-Bermejo B. and Fair\'en V. 1998 {\it J. Math. Phys.}
{\bf 39} 6162, and references therein.
\bibitem{Gumral} G\"umral H. and Nutku Y. 1993 {\it J. Math. Phys.} {\bf 34} 5691.
\bibitem{Plank1} Plank M. 1995 {\it J. Math. Phys.} {\bf 36} 3520.
\bibitem{Plank2} Plank M. 1995 {\it Helv. Phys. Acta} {\bf 68} 518.
\bibitem{Bermejo2} Hern\'andez-Bermejo B. and Fair\'en V. 2000 {\it Phys. Lett. A} {\bf
271} 258.
\bibitem{Perlick} Perlick V. 1992 {\it J. Math. Phys} {\bf 33} 599.
\bibitem{Hojman} Hojman S. A. 1991 {\it J. Phys. A: Math. Gen.} {\bf 24} L249.
\bibitem{Goedert2} Goedert J., Haas F., Hua D., Feix M. R. and Cair\'o L. 1994 {\it
J. Phys. A: Math. Gen.} {\bf 27} 6495.
\bibitem{Haas5} Haas F. and Goedert J. 1995 {\it Phys. Lett. A} {\bf 199} 173.
\bibitem{ath} Athorne C. 1991 {\it Phys. Lett. A} {\bf 159} 375.
\bibitem{Bermejo3} Hern\'andez-Bermejo B. and Fair\'en V. 1998 {\it Phys. Lett. A}
{\bf 241} 148.
\bibitem{Bermejo4} Yudichak T. W., Hern\'andez-Bermejo B. and Morrison P. J. 1998
{\it Phys. Lett. A} {\bf 260} 475.
\bibitem{Evans} Evans N. W 1990 {\it Phys. Rev. A} {\bf 41} 5666.
\bibitem{athor} Athorne C. 1992 {\it J. Diff. Equations} {\bf 100} 82.
\bibitem{att} Athorne C. 1991 {\it J. Phys. A: Math. Gen.} {\bf 24} L1385.
\end{thebibliography}
 \end{document}